\documentclass[acmsmall]{acmart}
\usepackage{subcaption}
\usepackage{multirow}
\usepackage{makecell}
\usepackage[utf8]{inputenc}

\newenvironment{aiquote}
  {\begin{list}{}{\setlength{\leftmargin}{0.7in}\setlength{\rightmargin}{0.6in}}\item[]}
  {\end{list}}

\newcommand\customquote[1]{\noindent``\textit{#1}''}

\newcommand{\totalReviews}{508,192}
\newcommand{\totalPCGReviews}{341,447}
\newcommand{\totalGenAIReviews}{166,745}
\newcommand{\totalGames}{158,899}
\newcommand{\totalPCGGames}{5,186}
\newcommand{\totalGenAIGames}{5,970}

\setcopyright{cc}
\setcctype{by}
\acmJournal{PACMHCI}
\acmYear{2026} \acmVolume{10} \acmNumber{7} \acmArticle{GAMES056}
\acmMonth{11} \acmDOI{10.1145/3831347}

\begin{document}
\title{Player Perceptions of Generative AI in Games: A Steam Review Analysis}

\author{Mahsa Bazzaz}
\affiliation{%
  \institution{Northeastern University}
  \city{Boston, Massachusetts}
  \country{USA}}
\email{bazzaz.ma@northeastern.edu}
\orcid{0009-0004-0022-9611}

\author{Seth Cooper}
\affiliation{%
  \institution{Northeastern University}
  \city{Boston, Massachusetts}
  \country{USA}}
\email{se.cooper@northeastern.edu}
\orcid{0000-0003-4504-0877}

%

\begin{abstract}
The rapid adoption of generative AI in game development has created large discussions among players, yet little empirical work has examined how players actually perceive AI-generated content. Employing quantitative methods, we study the adoption of generative AI in games in the Steam marketplace, using procedural content generation (PCG) as a baseline of a generative technology that was successfully integrated into games over several decades. Furthermore, using qualitative methods, we study player reception of generative AI by analyzing \totalReviews{} English-language reviews. We found that games disclosing generative AI use receive lower recommendation rates and more negative overall sentiment than PCG games. Thematic analysis of $600$ reviews shows that players perceive the use of generative AI in games as low developer investment in the game. Drawing on human-centered AI frameworks, we argue that successful generative AI adoption requires deploying generative AI for what players need, not for what makes development cheaper.
\end{abstract}



\maketitle
\section{Introduction}


Since the origins of role-playing games in the 1970s, tabletop game designers have used dice rolls, random encounter tables, and card draws to generate emergent gameplay~\citep{smith_analog_2015}. In video games, algorithmic content generation goes back to early roguelikes like Rogue (1980)~\citep{GAME_rogue} and space exploration games like Elite (1984)~\citep{GAME_elite}, which procedurally created vast, varied game worlds within severe memory constraints. These systems are now formally known as Procedural Content Generation (PCG), and are defined as algorithmic creation of game content with limited or indirect designer input~\citep{shaker_procedural_2016}.

Yet as Norman claims, ``for any new technology, only time is the key to user adoption and acceptance''~\citep{norman_design_2002}. When No Man’s Sky~\citep{GAME_nomanssky} launched in 2016 with 18 quintillion procedurally generated planets~\citep{parkin_no_2015}, players initially criticized its repetitive, meaningless landscapes and ``procedural generation'' became synonymous with ``empty'' and ``soulless'' content~\citep{ruffino_19_2024}. Through years of developers’ work showing craft and commitment, No Man’s Sky rehabilitated both its reputation and the broader perception of procedural generation, winning Best Ongoing Game for the second time at The Game Awards 2025~\citep{woods_every_2025}. Today, PCG is not just accepted but expected in roguelike games, and tags like ``procedural generation'', ``roguelike'', and ``randomly generated'' not only signal replayability but are the core identity of this genre.

More recently, the rapid adoption of ``generative AI'' (machine learning systems that produce content by learning patterns from training data) in game development has created a huge impact that was immediate and controversial. So much so, that in January 2024 (only two years after the first LLM-based chatbot release~\citep{wikipedia_large_2023}), Steam implemented a mandatory disclosure policy requiring developers to report AI-generated content~\citep{steam_news_ai_policy}. As of February 2026, more than 10,000 games have AI disclosure labels~\citep{steamdb_ai_content_disclose}. This is a massive adoption rate for a technology that was barely present three years prior.

This contrast between PCG's gradual and generative AI’s fast adoption presents a natural comparison case, but one that requires a careful caveat. Both PCG and generative AI are automated content generation systems that use computational processes to create game content~\citep{alharthi_generative_2025, bazzaz_playing_2026}. However, the differences in the nature of these generation systems have triggered very different player technology receptions. Therefore, our study does not compare PCG and generative AI as equivalent technologies. Instead, this study compares two `perception regimes' in which the difference in reception follows from how players perceive each technology, not from the technologies themselves. This creates a critical challenge for game developers, platform holders, and the 120 million active Steam users~\citep{clement_number_2023} navigating an increasingly AI-augmented marketplace. 

This paper provides the first systematic comparison of player reception of procedural generation versus generative AI in commercial games. Games have public review systems and engaged player communities, which provide an ideal testbed for studying these receptions. We collected \totalReviews{} reviews from games released between 2010--2025 on Steam, comparing games explicitly tagged with ``procedural generation'' (PCG, n=\totalPCGReviews{} reviews across \totalPCGGames{} games) to games carrying Steam's ``AI Generated Content Disclosed'' label (genAI, n=\totalGenAIReviews{} reviews across \totalGenAIGames{} games).
Our mixed-methods approach combines computational analysis with human-coded thematic analysis to answer 3 research questions:

\begin{itemize}
    \item \textbf{RQ1}: How has generative AI adoption evolved on Steam, and how does it compare to PCG in terms of market characteristics?
    \item \textbf{RQ2}: How different is player reception of generative AI compared to PCG, and what contextual factors moderate this gap?
    \item \textbf{RQ3}: What do players mention when they discuss generative AI in their reviews, and what themes explain the reception patterns found in RQ2?
\end{itemize}
This paper aims to measure how and why player reception differs and what themes structure these differences. As generative AI changes all creative industries, understanding how audiences perceive and evaluate AI-generated content is essential for responsible technology adoption that does not erode trust of consumers.

\section{Related Work}\label{sec:related}
We structure our prior work in three research areas: (1) procedural content generation as a generative technology that was successfully adopted into games (2) the emerging generative AI usage in commercial games and public perceptions of this, and (3) the broader methodological tradition of analyzing Steam reviews. These three areas contextualize both the technologies we compare (PCG and generative AI), the stakeholders whose perspectives we study (players and developers), and the methodological approach we build on (Steam review analysis). 

\subsection{Procedural Content Generation}
Procedural content generation (PCG) serves as the established baseline in our comparison as it is a generative technology that has been widely accepted by the game market over several decades. We draw on the PCG literature to define what PCG is, how it functions as an automation technology in games, and why it provides the right comparison case for understanding generative AI reception.

In the \citet{shaker_procedural_2016} definition of PCG, the ``content'' being generated can be levels, maps, textures, stories, items, quests, music, weapons, characters, and game rules. \citet{smith_procedural_2017} defines PCG as the process of using an AI system to author aspects of a game that a human designer would typically be responsible for creating, and emphasizing that the creator of a PCG system must capture some aspect of a designer's expertise. They further trace PCG's roots back to analog games, showing that the core motivations of replayability, surprise, and variety have persisted across decades and media~\citep{smith_analog_2015}. The motivations for PCG have always included reducing development costs by replacing some manual authoring with algorithms, augmenting human creativity so that small teams can produce content-rich games, enabling new game types with theoretically infinite content, and supporting player-adaptive experiences \citep{togelius_procedural_2013, shaker_procedural_2016}. 

What distinguishes PCG from other cost-reduction technologies is how it achieved acceptance: not by hiding its presence but by becoming inseparable from the gameplay experience itself. \citet{cook_game_2025} argues that all game design is fundamentally generative design and designing a game is basically to design an unpredictable and dynamic experience. Cook introduces the term \textit{procedural gameplay system} to describe generative systems that create content in ways players cannot predict and that change between play sessions. Under this framing, PCG is not an external production tool added to a finished design; it is an expression of design intent that produces input randomness in card games, emergent strategy in roguelikes, and unpredictable encounters in open-world RPGs. PCG, in other words, grew \textit{from within} game design practice rather than being imposed on top of it. Cook's survey data supports this: when players were asked to name three words they associate with procedural generation, the most common responses from both players and developers were "random," "roguelike," "infinite," and "replayability";  all terms describing gameplay experience, not production methodology \citep{cook_game_2025}. This integration into gameplay identity is what we mean when we say PCG is accepted: it is not merely tolerated but expected and valued as a design feature in genres like roguelikes and survival games.

We adopt this framing of PCG as an accepted automation technology in games for our comparison with generative AI. Both technologies use computational processes to produce game content that would otherwise require human labor, and both have been motivated by the same core goals of reducing development costs, enabling small teams to produce content-rich games, and creating novel player experiences. We note that by broad definitions, generative AI could be considered a subset of PCG. However, in both industry practice and player discourse, procedural generation and generative AI are treated as distinct categories, and Steam maintains separate tags and disclosure systems for each.

On Steam, PCG is primarily encountered through play. Players experience replayability, emergent encounters, and generated worlds. Generative AI, by contrast, is primarily encountered through observation and evaluation, whether via Steam's disclosure labels or players' own detection of AI-generated assets. This difference in legibility is what motivates our comparative framing: rather than treating PCG and generative AI as equivalent technologies, we compare them as two distinct reception regimes in which the same underlying question of ``how do players respond to computationally generated content?'' plays out through fundamentally different channels of player perception.

\subsection{Generative AI and Games}
Human Computer Interaction (HCI) research has examined how game developers perceive and adopt generative AI tools \citep{alharthi_generative_2025, panchanadikar_new_2024, boucher_resistance_2024, panchanadikar_solo_2024, vimpari_adapt_2023}. Literature on this topic shows developers' mixed views toward generative AI. 

In a study by \citet{vimpari_adapt_2023}, game professionals described this situation as an ``adapt-or-die'' scenario. They expressed the urgency and inevitability they felt around AI adoption, while also expressing concerns about job displacement and copyright. \citet{boucher_resistance_2024} extended this line of inquiry to early-career developers finding that ethical commitments to fellow developers and the future of their profession were as important as their practical concerns about tool usability and efficacy.

Focusing specifically on independent developers, \citet{panchanadikar_solo_2024} analyzed 3,091 online posts from indie game development communities and found that generative AI both promotes and threatens indie developers' efforts to innovate game production. In a related study, \citet{panchanadikar_new_2024} focused specifically on indie developers and found mixed perceptions, with some of them viewing generative AI as a ``new golden era'' for accessible game creation and others finding AI-generated assets as low-quality ``slap comps.'' A study by \citet{alharthi_generative_2025} confirms that although generative AI accelerates ideation and automates repetitive tasks, it raises persistent concerns about the originality of outputs and the potential undermining of human-authored content.

Beyond developer perceptions, recent work has examined how developers communicate their AI use to players. \citet{zhang_textual_2025} conducted a textual analysis of 584 AI disclosure statements on Steam from games released in 2024. Using BERTopic for distant reading and close reading of selected disclosures, Zhang identified the most commonly disclosed areas of AI application: voice generation, visual asset creation, translation and localization, and development workflow support. This work also found substantial inconsistency in how developers describe their AI use. The disclosures ranged from single vague phrases to detailed reflective explanations, highlighting significant gaps in transparency and standardization.

These developer-focused studies provide essential context for our work. However, while this prior work captures developer perspectives (from their attitudes toward AI adoption to how they disclose its use) our study provides the complementary player perspective, examining how players react to the same ecosystem.

\subsection{Steam Reviews as a Data Source}
Our study builds on a rich methodological tradition of using Steam reviews to understand player experience and behavior. Previous work has established Steam reviews as a valuable source of realistic player feedback, applied a range of analytical methods to extract insights from them, and demonstrated that review characteristics vary meaningfully across game types and player demographics \citep{lin_empirical_2019, wang_components_2020, busurkina_game_2020, petrovskaya_prevalence_2022}.
Early work of \citet{zagal_natural_2012} provided an overview of how natural language processing techniques can be applied to game studies research. Later, \citet{eberhard_investigating_2018} investigated the helpfulness of Steam reviews, finding that review length and time spent playing strongly influenced perceived helpfulness. \citet{Cody_identifying_2021} also studies steam reviews and showed that reviews contain rich affective data by using content analysis to identify games with therapeutic potential from player-reported experiences of coping, emotional regulation, and social connectedness.

Recently, \citet{guzsvinecz_length_2023} analyzed over 35 million reviews across 11 top-level video game genres, finding that emotional valence and sentiment differ between genres and that negative reviews tend to be longer than positive ones. This pattern is consistent with the richer diagnostic information in negative reviews that motivated our purposive oversampling strategy. Other studies have examined specific aspects of review behavior.

Our work differs from this prior literature in two key aspects. While previous Steam review studies have examined general player experience, we focus on player perception of generative AI specifically, treating the method of content creation as the object of study.

\vspace{\baselineskip}\noindent
Altogether, we position the paper's contribution relative to prior work as follows: while PCG's technical and player-centered motivations are well-studied, and while developer perspectives on generative AI adoption have been widely studied, the player perspective on generative AI in commercial games remains largely unexplored. Prior Steam review studies have developed the methodological tools to study player experience at scale, but have not applied them to generative technology reception specifically.
\section{Methodology}
\subsection{Data Selection}\label{sec:data}

We collected data using Steam’s public APIs and web scraping techniques. We collected a list of all applications on Steam via the ISteamApps API endpoint.\footnote{\url{http://api.steampowered.com/ISteamApps/GetAppList/v0002}} Then for each application, we obtained detailed metadata (like genre, price, release date, early access status) through the Steam Store API.\footnote{\url{http://store.steampowered.com/api/appdetails?appids=appid}}.

Since tag information and AI disclosure data are not available through official APIs, we used Python and BeautifulSoup to scrape and extract these values from individual store pages. We then collected English-language reviews for each application using Steam’s review API endpoint.\footnote{\url{https://store.steampowered.com/appreviews/appid}}.

\begin{table*}[!h]
\centering
\small
\caption{Descriptive characteristics across All Steam, PCG, and generative AI games. 
Game-level rows use the full app catalog (N=$\totalGames$); 
review-level rows use the collected PCG and Gen AI review corpus.}
\label{tab:descriptive}
\begin{tabular}{lrrrr}
\hline
\multicolumn{4}{l}{\textbf{Game-level characteristics}} \\
\hline
Characteristic & All Steam & PCG & Gen AI \\
\hline
N games & $\totalGames$ & $\totalPCGGames$ & $\totalGenAIGames$ \\
\hline
\multicolumn{4}{l}{\textit{Pricing}} \\
\quad Free-to-play (\%) & 23.3 & 12.5 & 22.4  \\
\quad Price, paid only (mean) & \$11.41 & \$14.70 & \$10.23\\
\hline
\multicolumn{4}{l}{\textit{Genre distribution (\%)}} \\
\quad Action & 31.7 & 51.2 & 31.1  \\
\quad Casual & 12.7 & 13.4 & 28.7  \\
\quad Adventure & 13.4 & 11.2 & 19.5  \\
\quad Indie & 8.9 & 14.4 & 11.0  \\
\quad Simulation & 4.2 & 2.6 & 3.1  \\
\quad Other & 29.1 & 7.2 & 6.5  \\
\hline
\multicolumn{4}{l}{\textbf{Review-level characteristics}} \\
\hline
N reviews &  & $\totalPCGReviews$ & $\totalGenAIReviews$   \\
\hline
\multicolumn{4}{l}{\textit{Release status}} \\
\quad Early Access reviews (\%) &   & 18.0 & 28.2  \\
\quad Received free copy (\%) &   & 3.5 & 4.0 \\
\hline
\multicolumn{4}{l}{\textit{Playtime}} \\
\quad Playtime at review, min (mean) &  & 1,276 & 1,360 \\
\quad Playtime at review, min (median) &   & 370 & 148  \\
\hline
\end{tabular}
\end{table*}

Steam's API returns content descriptor information only as numeric codes without labels. We obtained the mapping between content descriptor IDs and their descriptive names (e.g., Violence, Nudity, Adult Content) from Fredericksen's documentation of Steam's content rating system.\footnote{\url{https://fredericksen.substack.com/p/steam-video-game-ratings-and-mature}}

To identify games utilizing PCG or generative AI, we referenced SteamDB’s curated tag categories. SteamDB\footnote{\url{https://steamdb.info/}} is a widely used third-party database that tracks and aggregates metadata from Steam's store and community platforms. Because SteamDB does not permit automated scraping, we manually extracted application IDs for games tagged with ``Procedural Generation\footnote{\url{https://steamdb.info/tag/5125/}}''~(n=\totalPCGGames) and ``AI Generated Content Disclosed\footnote{\url{https://steamdb.info/tag/1368160/}}''~(n=\totalGenAIGames). 


These two tags operate differently in ways that are important to note. The ``Procedural Generation'' tag is community-generated and is applied through player votes and suggestions. The ``AI Generated Content Disclosed'' tag, on the other hand, is developer-reported to comply with Steam's mandatory disclosure policy. Our study therefore examines not generative AI in games in the abstract, but the reception of games that players encounter as AI games, in other words, those that have been disclosed in Steam's transparency ecosystem. We argue that this is precisely the population that matters for understanding player perception, as these are the games 
actively shaping public attitudes toward generative technologies in games right now. Undisclosed AI use and untagged PCG use represent related but distinct phenomena that fall outside this scope, and we return to the implications of this in Section \ref{sec:RQ1}.

We organized the collected data as a relational database with nine interconnected tables. The app details table (n=\totalGames{} games) serves as the central table that contains game-level metadata including commercial attributes (price, release date), content ratings (age restrictions, mature content flags), and aggregated review statistics. The other five tables hold details of tags, categories, genres, platforms, and language support. The reviews table contains all collected English reviews (n=\totalReviews{} reviews) as well as temporal attributes, engagement metrics (votes, playtime), and the sentiment score. The author table links reviews to player profiles that provide experience indicators (games owned, total playtime). 

\subsection{Sentiment Analysis}
For sentiment analysis, we do not perform extensive preprocessing, since punctuation and symbols (like emojis) help achieve more truthful sentiment detection. Our light preprocessing protocol only includes whitespace normalization (collapsing multiple spaces) and URL removal to eliminate non-content artifacts. 

We employed DistilBERT~\cite{sanh_distilbert_2019} for this purpose. This transformer-based model (unlike typical lexicon-based tools) captures contextual meaning, handles negation, irony, and domain-specific language through learned representations. This is particularly helpful for game review text, where lexicon-based approaches have been shown to misclassify reviews due to gaming-specific jargon and sarcasm~\cite{viggiato_causes_2021}. Such transformer-based models outperform lexicon and classical Machine Learning models on Steam review sentiment classification, and reach accuracies of 91\% or higher compared to 65\%-70\% accuracies of lexicon tools~\cite{hadinata_sentiment_2024, fadhlurrahman_sentiment_2023}. The DistilBERT model produces a continuous confidence score of scale -1 to +1, that classifies input into two classes (\textit{Positive}: score $>$ 0, \textit{Negative}: score $\leq 0$).

\subsection{Thematic Analysis}
To investigate the negative and positive views on generative AI between players (RQ3) and to complement our large-scale quantitative findings, we conducted thematic analysis on a stratified sample of Steam reviews where players explicitly discussed generative AI.
\subsubsection{Sample Selection}
We focused on reviews where players directly referenced generative AI, as these represent players’ conscious comments on these topics. Starting from our full dataset of English-language reviews of games disclosing generative AI use (n=\totalGenAIReviews), we applied seven filtering criteria to ensure review quality and relevance:
\begin{enumerate}
    \item Excluded reviews with zero helpfulness votes
    \item Excluded duplicate review texts
    \item Excluded reviews from users who received free copies
    \item Excluded reviews with excessive special characters (>20\% of text) indicating potential spam
    \item Excluded reviews with lengths more than 2 standard deviations from the mean length, in both directions (indicating they are either uninformative or spam)
    \item Included only reviews that contain at least one of the keywords: \textit{ai, ai-generated, ai generated, ai-created, ai created, generative ai, ai disclaimer, disclosed, disclosing, disclosure}
    \item Sorted reviews by number of upvotes and selected one most upvoted review for each game to avoid over-representation of games with larger numbers of reviews
\end{enumerate}

This filtering resulted in $809$ reviews. From this pool, we conducted purposive weighted sampling~\citep{etikan_comparison_2016} of $600$ reviews with stratification across three dimensions: (1) recommendation status (recommended vs. not recommended, weighted 30/70), (2) early access status (early access vs. full release, weighted 20/80), and (3) playtime at time of review ($<$2 hours vs. $\geq$2 hours, weighted 50/50).

We selected these stratification dimensions based on our quantitative analysis (Section \ref{sec:RQ2}), which revealed the contextual moderators to PCG and generative AI reception. We also intentionally oversampled not-recommended reviews (70\% of our final sample) based on prior research demonstrating that negative reviews contain more diagnostic information and specific details than positive reviews, which tend toward generic praise \citep{mudambi_research_2010, cao_exploring_2011}. Of the $600$ reviews, $35$ were reserved for iterative pilot coding used during codebook development, and the remaining $565$ were divided equally among the three coders (approximately $188$ reviews each).

\subsubsection{Codebook Development and Inter-Rater Reliability (IRR)}
We conducted a codebook-based qualitative content analysis following the iterative codebook development process outlined by \citet{hruschka_reliability_2004} and \citet{mcdonald_reliability_2019}. This approach uses a structured, collaboratively developed codebook applied consistently across multiple coders. In this study, three coders performed the qualitative analysis, all with experience as both game players and game development researchers with particular expertise in procedural content generation systems. Coder 1 familiarized themselves with the data by reading around 300 reviews and creating an initial draft codebook. Then all three coders independently coded a set of common pilot reviews in increments of 10 reviews, meeting after each increment to discuss disagreements and refine code definitions. We calculated both pairwise percent agreement and Krippendorff’s alpha \citep{krippendorff_computing_2011} on the jointly coded reviews. Pairwise percent agreement was calculated as the proportion of items on which each pair of coders assigned identical codes. Krippendorff's alpha~\citep{castro_fast_2017} was computed over all three coders simultaneously, accounting for chance agreement. We continued this iterative process until we reached an appropriate IRR, which happened at the 4th iteration with pairwise percent agreement (between 94.1\% and 97.1\%) and $\alpha = 0.920$ (good reliability threshold~\citep{marzi_k_2024} $\alpha \geq 0.80$). Our final codebook included 5 themes and 9 codes. After that, we proceeded to single-coding of the remaining data with spot-checks to monitor disagreement. Each of the three coders independently coded approximately $188$ reviews from the full sample of $600$ reviews.
\section{Findings}\label{sec:findings}
We structure our findings in three parts. First, we characterize the existence of PCG and generative AI on Steam. We do so by examining temporal emergences, pricing, content characteristics, and player engagement patterns (\textbf{RQ1}). Next, we analyze reception patterns to establish that PCG, as the baseline of successful technology integration in the market, receives more positive reception than generative AI games. We also investigate the contextual moderators that amplify this reception gap (\textbf{RQ2}). These quantitative findings set the base for our thematic analysis (Section~\ref{sec:themes}) in which we investigate why players respond differently to generative AI (\textbf{RQ3}).

\begin{figure}[htbp]
    \centering
    \includegraphics[width=0.8\linewidth]{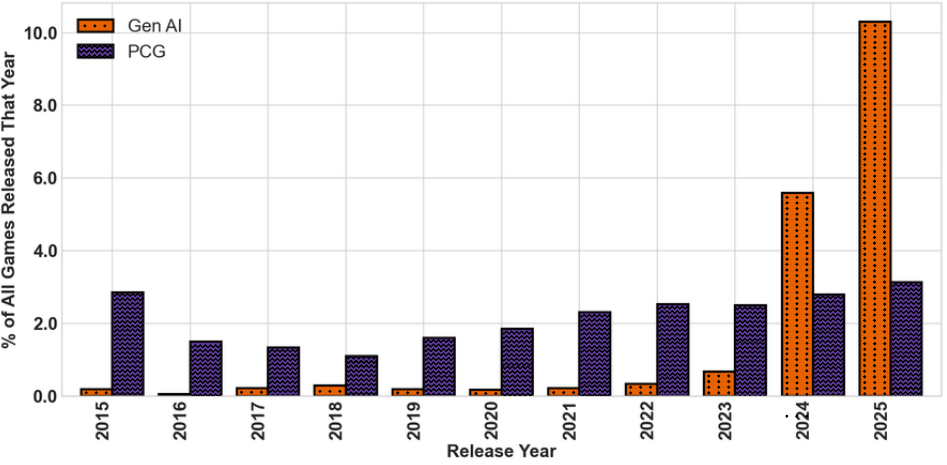}
    \caption{Percentages of game releases tagged as procedural generation (zig-zagged) or generative AI (dotted) per year.}
    \label{fig:temporal}
    \Description[Percentage of releases tagged PCG vs GenAI by year, 2015-2025]{A grouped bar chart showing the percentage of all games released each year from 2015 to 2025 that were tagged with Generative AI (orange, dotted) or Procedural Content Generation (purple, zig-zagged). PCG bars are relatively stable throughout, ranging from roughly 1.1\% to 3\% per year with a slight upward trend. GenAI bars are negligible from 2015 through 2022, then rise sharply: approximately 0.7\% in 2023, 5.6\% in 2024, and over 10\% in 2025, far exceeding the PCG proportion by that year.}

\end{figure}
\subsection{RQ1: PCG and Generative AI Games on Steam}\label{sec:RQ1}
Before comparing player perceptions, we first look into how PCG and generative AI games have different characteristics. These descriptive patterns reveal that generative AI represents a fundamentally different market segment than PCG.

The temporal patterns show that the video game industry has experienced a dramatic shift toward generative AI adoption in 2024--2025 (Figure \ref{fig:temporal}). Generative AI games were basically nonexistent (0.0--0.7\% of all Steam applications) from 2010--2023. In 2024, the number of games with AI-generated content jumped and accounted for 5.6\% of all Steam's releases. In 2025, this statistic increased again, comprising 10.3\% of all releases, representing over 3,000 disclosed AI games in just two years.

Table~\ref{tab:descriptive} presents descriptive characteristics of the full Steam catalog alongside the PCG and generative AI subsets. Significant market differences can be observed here. First, at the game level, generative AI games are significantly more likely to be free-to-play (22.4\% vs 12.5\%, $\chi^2$ = 186.40, p<0.001), and paid generative AI games are priced lower (mean \$10.23 vs \$14.70, Mann-Whitney U, p<0.001), suggesting that generative AI tools are disproportionately used to produce lower-budget titles. Second, PCG game reviews are typically posted after significantly longer playtime~(median 376 minutes vs 156 minutes, p<0.001). Third, genre distributions are also different. While both groups are dominated by Action games, games with generative AI disclosure are more present in Adventure (20.1\% vs 6.8\%) and Casual (14.9\% vs 6.6\%) genres, while games tagged as PCG are more present in Indie (19.7\% vs 6.4\%) and RPG (4.6\% vs 1.7\%) genres.

These differences may be confounding factors when comparing reception across the two groups. We therefore examine each as a contextual moderator of the perception gap in the following Section~\ref{sec:RQ2}, and later confirm their independent effects through multivariate regression.

As mentioned in Section \ref{sec:data}, ``Procedural Generation'' tag and ``AI Generated Content Disclosed'' tag come from different sources of information, and this asymmetry has implications for interpreting our findings. The ``Procedural Generation'' tag is community-generated and applied through player votes and suggestions. A game gets tagged as PCG when a considerable amount of players recognize procedural generation in that game.

The ``AI Generated Content Disclosed'' tag, on the other hand, is developer-reported to comply with Steam's mandatory disclosure policy. Since disclosure is self-reported, this tag likely undercounts the true population of AI-using games on the platform (games not disclosing their generative AI usage). Our thematic analysis (Section \ref{sec:themes}) documents this problem directly from the player perspective, where players report discovering undisclosed AI use through their own detection. 

The asymmetry in these two origins does not undermine the comparison for our purposes, because both tags represent how these technologies are publicly identified and encountered within the platform. We are studying player reception of games players perceive as PCG or AI games, and both tags are precisely how that perception is structured on Steam.

\subsection{RQ2: The Generative AI Reception Gap}\label{sec:RQ2}
Having characterized the distinct profiles of PCG and generative AI games, we now examine player reception through sentiment analysis of review text and the binary recommendation (thumbs up) system.
\subsubsection{Sentiment and Recommendation Patterns}\label{sec:recom_sentiment}
We examine two complementary signals of player perception: BERT-based text sentiment derived from review content, and the binary recommendation rate (Steam's thumbs-up system) as an explicit endorsement measure.

Interestingly, PCG games receive more positive sentiment in both text sentiment and binary recommendation, as compared to generative AI games.
As detailed in Table~\ref{tab:sentiment}, PCG reviews show higher positive (+16.0pp) and lower negative (-16.0pp) BERT sentiment compared to generative AI reviews. The 17.9 percentage point gap in recommendation rates is also notable, as it represents players' explicit willingness to endorse these games to others.

\begin{table*}[h]
\centering
\caption{Sentiment and recommendation percentages in PCG vs generative AI game reviews.}
\label{tab:sentiment}
\begin{tabular}{lrrr}
\hline
 & Negative\% & Positive\% & Recommendation Rate\% \\
\hline
PCG & 31.0 & 69.0 & 86.3 \\
Generative AI & 47.0 & 53.0 & 68.4 \\
Diff (PCG $-$ Generative AI, pp) & $-$16.0 & $+$16.0 & $+$17.9 \\
\hline
\end{tabular}
\end{table*}

A potential concern is that review-level sentiment differences could be driven by a few highly discussed games with extreme sentiment. To address this, we aggregated reviews to game-level means for games with $\geq$10 reviews (PCG: 2,056 games, generative AI: 1,752 games). The same pattern is still persistent~(Mann-Whitney U, p < 0.001). This confirms that sentiment differences are present because of patterns across many games rather than being driven only by a few outliers.

We ran the same analysis on the \textbf{AI-aware reviewers} (the subsection of the reviews who explicitly name generative AI) to observe the characteristics of reviews that are included in our thematic analysis \ref{tab:ai_aware}. These reviews (n=7,834) create 4.7\% of all reviews (n=166,515) and interestingly, are substantially more negative than the generative AI pool as a whole (see Table \ref{tab:ai_aware}). This establishes that the emerged qualitative themes in Section \ref{tab:thematic} describe the articulated form of an attitude that exists more diffusely in the broader population.

\begin{table*}[h]
\centering
\caption{Sentiment and recommendation percentages in all generative AI reviews versus AI aware generative AI reviews.}
\label{tab:ai_aware}
\begin{tabular}{lrrr}
\hline
 & Negative\% & Positive\% & Recommendation Rate\% \\
\hline
AI aware & 63.7 & 36.3 & 49.5 \\
All & 47.0 & 53.0 & 68.4 \\
Diff (PCG $-$ Generative AI, pp) & $-$16.7 & $+$16.7 & $+$18.9 \\
\hline
\end{tabular}
\end{table*}
\subsubsection{Genre Interactions}\label{sec:genre_source}
Not all game genres benefit equally from procedural generation or generative AI. To investigate whether each technology performs better in specific genres, we look into recommendation rates within the top 10 genres by review volume (Table~\ref{tab:genre}). 

The largest gaps appear in Action (+27.5pp toward PCG), RPG (+22.1pp toward PCG), and Strategy (+18.6pp toward PCG). The gap reduces for Indie (+8.1pp toward PCG) and Casual (+6.4pp toward PCG) games, and nearly disappears for Adventure (+2.3pp toward PCG) and Simulation (+1.6pp toward PCG). free-to-play is the only genre where games disclosing generative AI outperform PCG games (-1.4pp toward PCG).

\begin{table*}[h]
\centering
\caption{Recommendation Rate Gap by Genre}
\label{tab:genre}
\begin{tabular}{lrrrr}
\hline
Genre & PCG & Generative AI & Gap (PCG-Gen AI, pp) \\
\hline
Action & 87.8 & 60.3 & +27.5 \\
RPG & 87.0 & 65.0 & +22.1 \\
Strategy & 86.2 & 67.5 & +18.6 \\
Massively Multiplayer & 77.8 & 61.3 & +16.5 \\
Indie & 86.1 & 78.0 & +8.1 \\
Casual & 84.8 & 78.4 & +6.4\\
Adventure & 82.8 & 80.5 & +2.3 \\
Simulation & 75.4 & 73.8 & +1.6 \\
Free-To-Play & 73.1 & 74.5 & -1.4 \\
\hline
\end{tabular}
\end{table*}
\subsubsection{Price Effect: Free vs Paid Games}\label{sec:price_effect}
To investigate whether the disproportionate amount of free-to-play games disclosing generative AI use (Section \ref{sec:RQ1}) explains the perception gap, we split games into free and paid subgroups and compared PCG vs. generative AI recommendation rate within each (Figure \ref{fig:price}).
In PCG games, the recommendation rate gap between paid and free-to-play is 3.8 percentage points (86.9\% vs 83.1\%). In generative AI games, however, the recommendation rate gap between paid and free-to-play goes up to 28.8 percentage points (78.9\% vs 50.1\%). This pattern suggests that a large part of the overall generative AI reception deficit is in the free-to-play games. The usage of generative AI in these low-investment games seems to be negatively evaluated by players.

\subsubsection{Early Access: A Protective Framing for Generative AI}\label{sec:ea_effect}
Descriptive statistics reveal that 28.2\% of generative AI reviews come from Early Access titles compared to 18.0\% of PCG reviews (Table ~\ref{tab:descriptive}) which raises the question whether the ``work in progress'' framing of Early Access moderates the reception gap.
Interestingly, Early Access indeed provides substantial protection for games disclosing generative AI (Figure \ref{fig:status}). Generative AI Early Access titles have an 77.8\% recommendation rate while PCG Early Access titles have a 85.8\% recommendation rate~(only a +8pp gap). However generative AI full-release games have a dramatically lower recommendation rate of 65.7\% compared to PCG full releases 86.8\%~(+21.1pp gap). 

\subsubsection{Playtime Effect: Quick Rejection vs Longer Engagement}\label{sec:playtime_effect}
PCG players invest significantly more time before posting their review than generative AI players (Mann-Whitney U, p < 0.001). The median PCG player spends 370 minutes (6.1 hours) at the time of writing the review, compared to only 148 minutes (2.4 hours) for generative AI reviewers.

\begin{figure}[h!]
\centering
\begin{subfigure}{0.32\textwidth}
    \centering
    \includegraphics[width=\linewidth]{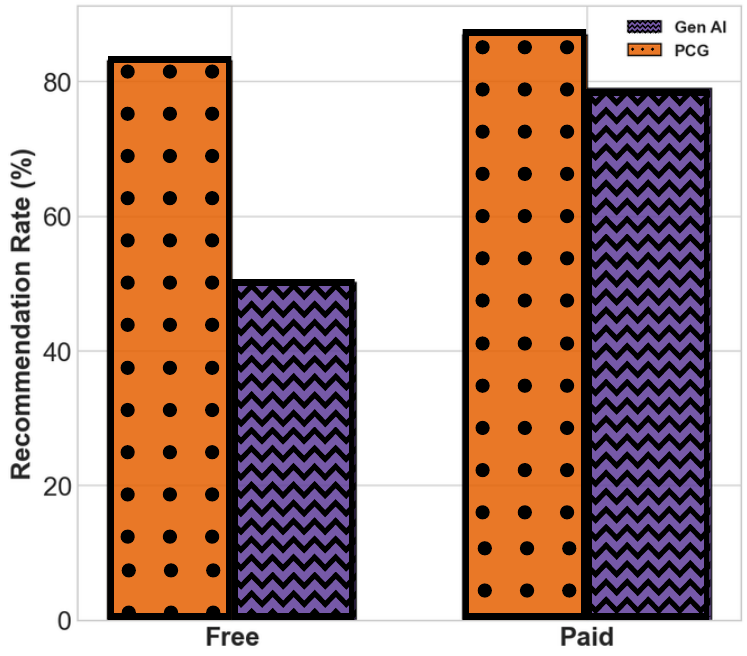}
    \caption{Price Effect}
    \Description{Under ``Free,'' PCG recommendation rate is 83.1\% 
    and Gen AI is 50.1\%, a 33pp gap. Under ``Paid,'' PCG is 86.9\% 
    and Gen AI is 78.9\%, an 8pp gap.}
    \label{fig:price}
\end{subfigure}
\hfill
\begin{subfigure}{0.32\textwidth}
    \centering
    \includegraphics[width=\linewidth]{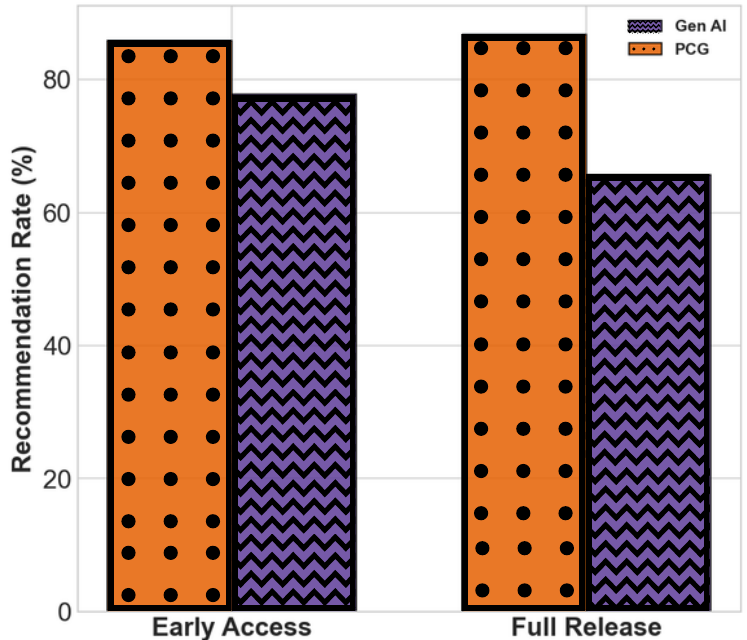}
    \caption{Early Access Effect}
    \Description{Under ``Early Access,'' PCG recommendation rate is 
    85.8\% and Gen AI is 77.8\%, an 8pp gap. Under ``Full Release,'' 
    PCG is 86.8\% and Gen AI is 65.7\%, a 21pp gap.}
    \label{fig:status}
\end{subfigure}
\hfill
\begin{subfigure}{0.32\textwidth}
    \centering
    \includegraphics[width=\linewidth]{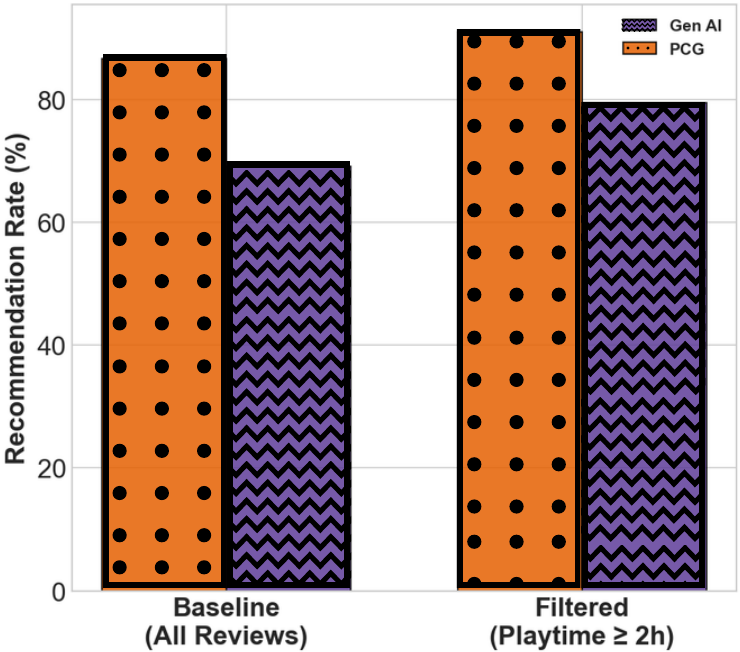}
    \caption{Playtime Effect}
    \Description{Under ``Baseline (All Reviews),'' PCG recommendation 
    rate is 86.7\% and Gen AI is 69.1\%, a 17.6pp gap. Under 
    ``Filtered (Playtime greater or equal 2h),'' PCG is 91.0\% and Gen AI 
    is 79.5\%, a gap that narrows to 11.5pp.}
    \label{fig:playtime}
\end{subfigure}
\caption{Recommendation rates (\%) for PCG (dotted) and generative AI (zig-zagged) games
across three contextual moderators: (a) price, (b) release 
status, and (c) reviewer playtime (at time of review). In all three conditions, 
PCG outperforms generative AI, but the gap varies in each. 
free-to-play games show the largest gap, while Early Access 
and higher playtime show narrower gaps.}
\Description{Three side-by-side bar charts comparing PCG (dotted bars) and generative AI (zig-zagged bars) recommendation rates across price, release status, and playtime.}
\label{fig:moderators}
\end{figure}

To investigate this pattern more, as seen in Figure \ref{fig:playtime}, we filtered out reviews submitted with less than 2 hours of gameplay (players who may have given up immediately within Steam's refund window). Filtering low-playtime reviews improves generative AI recommendation rate by +10.4pp (69.1\% to 79.5\%) but only +4.3pp for PCG (86.7\% to 91.0\%). This dramatic difference suggests that many negative generative AI reviews come from players who rejected the game (or the generative AI in the game) quickly, while negative PCG reviews come from more engaged players with longer experience.

Lastly, to confirm the reception deficit is not an artifact of correlated confounds (free-to-play status, Early Access, and playtime), we fit a logistic regression predicting positive recommendation controlling for all three factors plus a source $\times$ price interaction. Generative AI remains the strongest predictor of negative reception (OR = 0.63, 95\% CI [0.62, 0.64], p<.001), and the source $\times$ price interaction is particularly damaging to reception (OR = 0.36, p<.001), consistent with the descriptive gap reported in Section~\ref{sec:price_effect}.

Overall, the reception gap between PCG and generative AI is not uniform. As we saw, it varies dramatically by genre, greatly reduces in Early Access games, and reduces when filtering out low-playtime reviews. These contextual findings suggest that generative AI reception depends critically on \textit{where} it's used (genre), \textit{how} the technology is framed (early access vs full release and free vs paid), and \textit{who} engages with it (players with different playtime at time of review). Having established these quantitative patterns, we now turn to understanding \textit{why} players respond so differently to these two generative systems through thematic analysis of review content.

\subsection{RQ3: Player Perception on Generative AI}~\label{sec:themes}
Our thematic coding of 600 Steam reviews by three coders resulted in five distinct themes characterizing player attitudes toward AI-generated content. These qualitative findings help to clarify the sentiment patterns present in reviews.

\begin{table*}
\caption{Qualitative codebook and Thematic Analysis of player attitudes toward generative AI in Games. The table presents the 5 themes and 9 codes, with exemplar quotes from the coded reviews.}
\label{tab:thematic}
\begin{tabular}{p{2.9cm}p{2.2cm}p{8.8cm}}
\toprule
\textbf{Theme} & \textbf{Codes} & \textbf{Exemplar Quote} \\
\midrule

\textbf{T1: Perceived\newline Quality Deficit}
& 
\textit{ai\_quality\_bad} 
& \customquote{I find it baffling to use ``AI art" in a genre as low on graphical expectations as incremental/idle games. I would much rather play the game that's entirely your original creation, even if you personally find your own artistic capabilities lacking.}
\\
\cmidrule(lr){2-3}

& \textit{ai\_value\_bad} 
& \customquote{In fact the developers want ONE HUNDRED AND TWENTY SIX UNITED STATES DOLLARS from you... to buy all the AI slop and stolen public domain/stock photos they shoved into this.  See why I have a problem with this?}
\\
\cmidrule(lr){2-3}

& \textit{ai\_spillover}
& \customquote{You start thinking... If the art is AI, and the music is AI, and the script is possibly partially AI.... What is the point in even engaging with this? Now obviously there is a studio of real people behind this game, and obviously there are elements of it that are NOT AI. Obviously this still took time and effort to put together... But the use of AI should just be a supplement, something to make PART of a job a little easier.... When it becomes the majority of what you're seeing and hearing, it starts to feel less like a supplemental tool and more like a crutch.} \\
\cmidrule(lr){2-3}

& \textit{ai\_technical} 
& \customquote{OOS errors have been plaguing multiplayer for over a year now making the game entirely unplayable past year 40 if you're lucky to get this far. Clearly these devs are only using AI to write their code ever since 4.0 and that is very clear in their artwork as well.} \\

\midrule

\textbf{T2: Ideological \& Ethical Resistance} & 
\textit{ai\_stigma}
& \customquote{This hurts me to do because I used to LOVE this game. But the developer has turned to using AI and it sucks now.}\\
\cmidrule(lr){2-3}
 & \textit{ai\_ethics \newline}
& \customquote{If you can't pay artists you don't deserve my support. Not even worth the 6 dollars.} 

\\
\midrule

\textbf{T3: Acceptance} & 
\textit{ai\_good\_use} &
\customquote{This may not be the first game trying to let AI assist players in influencing the plot, it is the best one I have played so far that does so without letting me feel the use of AI is out of the place: after all, if the setting and the original story is about tale-telling, it is the most natural that we can get the AI to be the annoying and arrogant king who would roast the player for not being able to come up with a sufficiently interesting story!}\\
\midrule

\textbf{T4: Transparency \& Trust} & 
\textit{ai\_disclosure\_mismatch} &
\customquote{The developers describe how their game uses AI Generated Content like this: Some Audio/ music was generated by using AI Sorry, but I can see that there's more than the audio and I hate dishonesty.}

\\
\midrule





\textbf{T5: The Bar Keeps Rising} & 
\textit{should\_use\_ai} 
&
\customquote{Voice acting started off bad and just got worse and worse - constant mispronunciations and some sentences simply don't make sense in English. What I also don't get is the ``AI Companion" character voiced by TTS when we literally have real life AI models that can perfectly imitate real people without issue.} \\

\bottomrule
\end{tabular}
\end{table*}

\noindent$\bullet$~\textbf{T1: Perceived Quality Deficit: ``It Looks Like Nobody Cared''}
This theme is the dominant negative view to generative AI, capturing player perception that AI-generated content signals a lack of care and human investment. Players think of AI-generated content as ``rushed'', ``lazy'', and ``soulless'', and the presence of such content leads to an overall impression of low effort that extends beyond the assets themselves.

\begin{aiquote}
    \customquote{``wrong" looking AI slop instead of genuinely and sincerely created game assets.  It's hard to say if this was because the developer didn't have the skill to do the job of creating real game assets or couldn't afford to hire someone who does, but it also doesn't matter.  What matters is that this looks bad because of that decision.  AI slop sticks out like a sore thumb... and it comes across as insulting.}
\end{aiquote}

Players identified quality failures across multiple aspects. In visual assets, reviewers pointed to anatomical errors such as extra fingers and incoherent proportions. In AI-generated voice acting, they complained about robotic delivery of voices that broke immersion. In text, particularly translations, reviewers noted obvious errors that suggest no human proofreading of AI output.

\begin{aiquote}
    \customquote{(...) And my last complaint is that, I am not necessarily against using AI to assist in making a game, but someone has to go back and quality control it... make sure it all makes sense.  There are numerous problems in the artwork, from formless blob shapes to extra/spikey fingers to eye and tattoos changing between panels to anatomy doing things that would make me scream if it were mine... nobody touched up the AI generated art to make it make sense.}
\end{aiquote}

What unifies these responses is that AI-generated content is detected by players as low quality, and it signals that the developer did not invest the effort to review or refine the output.

This theme also extended into price perception and technical quality. Several reviews explicitly questioned the fairness of charging for AI-generated content.
Additionally, some reviewers connected game-breaking bugs to AI-assisted coding, a code we labeled \textit{ai\_technical}. While relatively rare, this code captures an emerging concern around ``vibe coding'' (writing software by describing desired functionality to an LLM~\citep{meske_vibe_2025}) and the practice of using LLMs to generate game code without sufficient developer oversight.

\begin{aiquote}
    \customquote{This is a badly done attempt to make an RTS game using AI slop and ``vibe coding" (when you don't know how to make a game but want to do it anyway, ChatGPT is there to make gaming worse for everyone).}
\end{aiquote}

Our code \textit{ai\_spillover} captures that extensive usage of AI poisons the evaluation of the entire product. Once players identify multiple assets as AI-generated, they begin to question the authenticity of everything, even elements that may have been human-made, undermining trust in the whole experience.

\begin{aiquote}
    \customquote{game is absolutely dripping with zero effort AI slop, to the point where many of the non-AI assets still feel like they were run through a plagiarism engine to shave off any personality or character. like, if the songs on the radio station aren't AI, they were absolutely written and recorded by a serial killer because that is how little humanity is on display.}
\end{aiquote}

Taken together this theme suggests that AI use triggers a reassessment of developer competence with quality concerns varying from aesthetics to pricing to technical performance.

\noindent$\bullet$~\textbf{T2: Ideological \& Ethical Resistance: ``No place for AI in games''}
While Theme T1 relates to players' concerns with what AI produces, Theme T2 captures players' objections to what AI represents. These reviews reject generative AI on principle, regardless of whether the output quality is acceptable. This distinction proved to be the most contested boundary in our coding process. When a reviewer writes ``AI slop'' without elaboration, coders diverged on whether this constitutes a quality judgment or an ideological rejection. Despite this ambiguity, a group of reviews made the principled stance explicit.

\begin{aiquote}
    \customquote{I LOVE THIS GAME, BUT I CANT MORALLY RECOMMEND IT FOR 1 SPECIFIC REASON THAT DOESN'T AFFECT GAMEPLAY BUT STILL SETS A BAD PRECEDENT: It uses AI artwork for some of the set dressing. (...) The game itself is so close to being a perfect 100/100 but the fact that it uses AI slop, makes me the consumer feel morally wrong for indulging in the game, because its setting the precedent that AI slop is okay to use. 
}
\end{aiquote}

This pattern (positive sentiment coupled with negative recommendation) provides qualitative grounding for the differences in recommendation and sentiment analysis observed quantitatively in Section \ref{sec:recom_sentiment}.

\begin{aiquote}
    \customquote{While the game itself was reasonably fun, I am ultimately going to put a thumbs down for the rampant use of AI `art'. Yes, it's a free game, but the principle remains the same: the more AI art is used in games without anyone fighting back, the more normalised the idea becomes and the more acceptable devs and publishers on the other end of the scale (looking at you, triple A industry) will think it is ok to screw artists over. While AI has its uses, it should be used responsibly - and definitely not in a way that devalues the work of real living human beings.}
\end{aiquote}

Even though players believe the game is fun, they choose to downvote anyway as a form of consumer activism. The \textit{ai\_stigma} code captures this rejection, including instances of retroactive rejection (where enjoyment ceases upon discovering AI use) and conditional recommendations (``would change my review if AI were removed'').

Ethical concerns centered on artist displacement and the normalization of AI use across the industry. Several reviews framed their negative recommendation as a deliberate act of resistance, which is not against the individual game, but against the trajectory they perceive the industry taking.

Other ethical aspects included concerns about training data (stolen content), the flooding of Steam with low-quality AI-generated games, and in a small number of cases environmental impacts of generative AI. Notably, many reviews in this theme expressed willingness to even accept lower-quality human-made assets over highly polished AI-generated alternatives, suggesting that for these players, the process of creation matters more than the output.

\noindent$\bullet$~\textbf{T3: Acceptance: ``AI Done Right''}
Despite the overwhelmingly negative reception of generative AI in our sample, a number of reviews expressed acceptance and enjoyment toward generative AI. This theme reveals the conditions under which AI use is welcomed, providing a counterpoint to T1 and T2. 

The most common condition was proportionate use of AI in areas where it adds genuine value or novelty. Reviews liked developers who used AI as a tool rather than as a wholesale substitute for creative labor.

\begin{aiquote}
    \customquote{The developer used AI to upscale the original artwork... we all know AI slop is disgusting but this seems like one of those really rare cases that it's okay, because the developers didn't use AI as a substitute for real, genuinely created game assets, just rizzing up the original game assets.}
\end{aiquote}

AI as a novel gameplay mechanic was also enjoyed among players. When AI was the game’s premise (such as LLM-driven dialogue or procedurally generated narratives) players evaluated it on its interactive merits rather than as a cost-cutting measure.
\begin{aiquote}
    \customquote{develepers [\emph{sic}] of this game are definitely pioneers of AI-driven games genre. Believe me or not - I loved possibility to hold dialogs with AI in this game!}
\end{aiquote}

Early access status also moderated reactions, consistent with the quantitative findings in Section \ref{sec:ea_effect}. Some reviewers explicitly extended the ``benefit of the doubt'' to early access titles, treating AI assets as acceptable placeholders under the assumption that they would be replaced before full release.
\begin{aiquote}
    \customquote{
    This game uses AI art assets. It is in early access, and I am giving the developer the benefit of the doubt that they are ``placeholders" while they look for a ``real artist" to do the art for their video game they supposedly care about.}
\end{aiquote}

\noindent$\bullet$~\textbf{T4: Transparency \& Trust: ``They Tried to Hide It''}
This theme captures the erosion of player-developer trust when the stated extent of AI use does not match what players observe. Since Steam’s AI disclosures are self-reported by developers, players have become active investigators, cross-referencing disclosure statements against observable evidence in the game. In some cases, developers either failed to disclose AI use or did so only after being caught.

\begin{aiquote}
    \customquote{Developer claims they have always been transparent about AI-generated content, but a simple look at the Wayback Machine makes it clear that's absolutely untrue.}
\end{aiquote}

It was also seen in the form of understatement, where the disclosure technically acknowledged AI use but minimized its scope in ways players found misleading. This mismatch is trust-damaging and it undermines the disclosure system as a trust-building mechanism.

\begin{aiquote}
    \customquote{It says that it uses AI to ``generate some pictures as creative references and sources of inspiration", but when you get hands with 3 fingers, or more then five, you can tell that it is used for more then just ``reference and inspiration."}
\end{aiquote}

This theme suggests that the disclosure mechanism may actually increase distrust between players and developers when players feel that disclosures are missing, incomplete, or misleading and contrary to their playing experience.

\noindent$\bullet$~\textbf{T5: The Bar Keeps Rising: ``Why Didn't They Use It?''}
Lastly, a small but notable theme contains reviews where players criticize games for not using generative technologies and
AI-assisted features, particularly in translation and voice synthesis, where some reviewers criticized games for having poor translations when AI tools were readily available.

\begin{aiquote}
    \customquote{The bad and lazy translations, while they don't get in the way of gameplay, are still an eyesore, and a good representation of the game's quality. Come on, we live in the age of AI - the developer could have fixed all bad translation in the game within 2 hours.}
\end{aiquote}

This theme provides a counternarrative to Themes T1, T2 and T4: while some players reject these technologies (either based on quality or principle), others object to their absence. 

\section{Discussion}
\subsection{From Quality Failures to Perceived Intent}
\citet{bazzaz_playing_2026} conducted a study focused on the effects of player beliefs about whether generative-AI was used in a game level on how they experience that level. Their study involved players playing short Super Mario Bros. and Sokoban levels without knowing the source of creation of the levels. The study showed that when players are asked to distinguish between AI-generated and human-created levels they cannot reliably identify each, and their classification accuracy is barely above chance. But interestingly, they also report that players hold more negative attitudes toward generative AI than toward PCG, and content believed to be AI-made was rated as less fun and more frustrating than identical content believed to be human-made. 

Our thematic analysis documents the same detection-then-judgment sequence in the setting of a large marketplace (Steam). Players who encounter games on Steam are, like \citet{bazzaz_playing_2026}'s participants,  attempting to form their own beliefs about games. Whenever they observe any quality failures, those function as the detection cues that trigger the negative bias that study observed. We also observed that our participants had the same ethical and ideological concerns against generative AI that they observed in their participants. Our Theme T2 (Ideological and Ethical Resistance) includes some players who explicitly acknowledge a game's enjoyment but leave a negative recommendation, framing this as deliberate consumer activism against the normalization of AI in creative industries.

In \citet{bazzaz_playing_2026}'s study, the negative attitude manifests as lower fun and higher frustration ratings. In our corpus of \totalReviews{} reviews, the 17.9 percentage point recommendation gap 
between PCG and generative AI games represent the same negative player experience measures.

\subsection{The Transparency Dilemma}
The reception examined so far is a response to generative AI itself. Players evaluate the technology's presence and read it as a signal about the developer, independent of how that presence is communicated. On Steam, however, AI use is also surfaced through a disclosure label, and a separate body of work shows that the act of disclosure carries trust consequences of its own. \citet{schilke_transparency_2025}, across thirteen experiments, find that actors who disclose using AI for a task are trusted less than those who do not. This distinguishes a reaction to the technology from a reaction to its disclosure.

Steam's AI disclosure requirement was designed to build trust by making AI use visible to players. However, this transparency mechanism \textit{itself} has become a controversy. Developers complain that the disclosure label alone triggers negative reviews regardless of content quality or how minimally generative AI was used. Epic Games CEO Tim Sweeney publicly criticized Steam’s approach. He argued that AI disclosures ``make no sense'' because generative tools will inevitably become ``involved in nearly all future production,'' and that highlighting generative AI use in games is as arbitrary as disclosing which game engine or programming language was used~\citep{down_epic_2025}. The backlash caused Steam to quietly revise the disclosure policy in January 2026~\citep{chalk_steam_2026}. The updated policy narrows the mandatory disclosure to generative AI used to create content that ships with the game, excluding any generative AI-powered development tools.

Our findings suggest that Steam's transparency system in its current form does not necessarily work as a trust-building mechanism. Our Theme T4 shows a pattern of players cross-referencing disclosure statements against observable in-game evidence and going as far as checking archived versions of developer pages via the Wayback Machine to treat any inconsistencies as evidence of intent to deceive.

\citet{schilke_transparency_2025} saw the same dynamic in their empirical study of supervisors, professors, analysts, and creative professionals. They found that third-party exposure of AI use by a different party, produces a larger trust penalty than self-disclosure, meaning that being caught using generative AI (without disclosing it) is significantly worse than disclosing it upfront. This mechanism aligns with what we observe in our T4. Developers who comply with mandatory AI disclosure but minimize the scope of how generative AI was used (e.g., disclosing audio generation while players can see AI artifacts in the visual assets) do not receive credit for partial transparency. Instead, they face the third-party-exposure penalty as \citet{schilke_transparency_2025} puts it.

\citet{zhang_textual_2025} pinpoints the upstream source of this problem. Their analysis of 584 Steam AI disclosures showed that disclosures range from single vague phrases to detailed reflective explanations, with no platform-enforced standards for scope, format, or specificity. This means that the disclosure system creates the expectation of transparency but permits practices that fall short of it. As \mbox{\citet{zhang_textual_2025}} notes, this risks what \citet{floridi_translating_2019} calls ``ethics bluewashing'', i.e, creating ethical guidelines to only create a false image of social responsibility. We believe players have developed sophisticated detection practices precisely because the Steam disclosure system has not closed the information gap it was designed to address.

The implication is not that disclosure is counterproductive, but that disclosure systems require an intentional design that enforces which assets were AI-generated, what human curation was applied, and why generative AI was used. This would be a shift toward the full transparency which we believe is the only durable path to trust.

\subsection{Procedural Generation as Generative Design}
\citet{cook_game_2025} argues that game design is fundamentally generative design. He states that PCG is not a production tool that is added on top of a finished design, but it's an expression of design intent that creates emergent gameplay. This distinction maps directly onto what we observe in player perception of generative AI. In Theme T1, players encounter quality failures in AI-generated assets directly during play, but their complaints in many cases do not stop at the quality and they extend that to evidence that the developer may have not invest the care that the product needed. In Theme T2, players reject AI use on principled grounds regardless of whether the output is acceptable. Both themes center on what AI usage \textit{signals} about the developer's relationship to their craft and to the community, not on whether the technology creates the experience that Cook identifies as the basis for PCG's acceptance.

This asymmetry has implications for how developers deploy generative AI. \citet{cook_game_2025}  notes that the game industry's understanding of procedural generation has crystallized around a few specific use-cases, restricting innovation and pigeonholing its use-cases to a handful of repetitive and commercially-focused applications. Our data suggest that generative AI is crystallizing around replacing human labor in asset production (art, voice, translation) rather than creating novel gameplay experiences. This is an even narrower and more commercially-focused application pattern of PCG that Cook mentions.

The cases of AI acceptance we identified in our thematic analysis (LLM-driven dialogue systems, AI-powered narrative generation, dynamic content that responds to player input) are precisely the cases where AI functions as what \citet{cook_game_2025} would call a procedural gameplay system that shapes player experience rather than merely populating it with cheaper assets. \mbox{\citet{yannakakis_experience_2011}} formalized this principle for PCG decades before generative AI was popularized. They argued that the quality of computationally generated content should be measured by player experience, not by artifact quality metrics. The generative AI adoption trajectory we observe on Steam suggests that this lesson has not yet been applied. The technology arrived from an efficiency-optimization perspective instead of a player-experience view, and players are responding to that difference.

\subsection{Toward Human-Centered AI in Games}
Beyond platform policy, our findings also point toward a fundamental design orientation problem. The majority of generative AI applications that players reject share a common characteristic: they are \textit{technology-centered} rather than \textit{human-centered}, deploying AI based on what it can automate rather than on what players need. \citet{norman_design_2002} articulated the foundational principle that good design must start from the needs and capabilities of the user, not from the affordances of the technology. When technology drives design rather than the reverse, the result is what Norman calls a mismatch between the designer's conceptual model and the user's, producing frustration, distrust, and rejection. The sentiment patterns we observe suggest exactly this kind of mismatch. Developers adopt generative AI because it reduces production costs or compensates for missing skills; players evaluate the resulting product against their expectations of creative investment and craftsmanship.

\citet{riedl_human_2019} defines human-centered AI as requiring two complementary capabilities: AI systems that understand humans from a sociocultural perspective, and AI systems that help humans understand them. Both capabilities map onto failure modes visible in our data. The first capability—AI understanding its human context—connects to what we observe in the quality deficit and spillover codes (T1). Players reject AI-generated content not merely because it contains technical errors (extra fingers, robotic voice delivery, incoherent translations) but because those errors signal that the developer did not account for how players read and interpret game assets as evidence of care. AI-generated content that is ``detectable'' fails precisely because the system, and the developer deploying it, lacks what Riedl calls a theory of mind about players: an understanding of how they will perceive, evaluate, and emotionally respond to the content they encounter. The second capability—helping humans understand AI—connects directly to our transparency and trust theme. When intelligent systems inevitably make mistakes or violate expectations, Riedl argues that people need explanations and enough information to revise their mental model. Steam's disclosure system was designed to serve this function, yet our data reveal that it frequently achieves the opposite, transforming disclosure into a liability when developers understate the extent of their AI usage.

Taken together, these frameworks suggest that the path toward successful generative AI adoption in games is not primarily a matter of improving output quality, though that matters as well. It requires reorienting the design question from ``\textit{what can AI generate for us?}'' to ``\textit{what do players need that AI can uniquely provide?}'' We believe using generative AI as a cost-reduction tool for asset production will continue to perpetuate the perception gap we observe. On the other hand, using generative AI as a player-driven system has the potential of receiving a fundamentally different and positive perception by players. In our Themes 3 and 5, we observed cases where generative AI is currently accepted (dynamic narrative systems, responsive dialogues, AI-driven content that evolves with the player) and they all share the property that they serve some player need that human-authored content cannot replicate at the same scale. These are applications of AI that grow from within the logic of game design, analogous to how PCG grew from the desire for replayability and emergent gameplay. For game developers, this is indeed a path forward.
\subsection{Limitations and future directions}
Our study has several limitations that we acknowledge in this section.
First, we only analyzed reviews written in English. Steam is a global platform with a substantial proportion of non-English reviews \citep{lin_empirical_2019}, and player attitudes toward generative AI may differ across linguistic and cultural contexts~\citep{wu_investigating_2020}---particularly given that ethical concerns around AI-generated content and artist displacement are shaped by culturally specific labor norms and creative values. Our findings should be interpreted as reflecting primarily English speaking populations, and cross-cultural replication of this study is an important extension for future work.

Second, we estimated the lifetime and temporal position of games using the release date as advertised on the Steam Store page. This date is an approximation because developers are permitted to change their listed release date. This may introduce noise into our temporal analyses, particularly the analysis of generative AI's rapid rise in 2024--2025 and the Early Access comparisons.

Third, our thematic analysis sample was drawn from reviews that explicitly mention generative AI through keyword filtering. This means our qualitative findings capture players' \textit{conscious} discourse about these technologies but necessarily exclude reviews where players react to generated content without naming it. The attitudes of players who notice quality issues but do not connect them to a generative technology represent an important gap that future work could address.

Fourth, our two comparison groups are identified through different mechanisms, community tagging for PCG and developer disclosure for generative AI. However, as established in  Section~\ref{sec:data}, both tags are visible components of 
the Steam marketplace that players encounter directly, and both represent how these technologies are publicly identified within the platform's own ecosystem. The difference in their origins is therefore consistent with our research question. What falls outside our scope is undisclosed AI use and unrecognized PCG use, but as of now, no platform-level alternative exists for constructing these groups. Future work could move toward capturing these populations by using independent detection methods to create more comparable sampling distributions.

Lastly, as player attitudes toward generative AI are evolving rapidly, we hope that future longitudinal studies would track sentiment trajectories as the technology usage matures in this industry and disclosure norms stabilize.

\section{Conclusion}
This paper compared player reception of procedural content generation and generative AI in commercial games through a mixed-methods analysis of Steam reviews. Our findings establish that the sentiment gap between these two generative technologies is persistent. Our thematic analysis reveals that the gap is not simply about output quality. Players evaluate generative AI through the lens of developer intent, questioning the reasoning, ethics, and transparency of its use. We argue that PCG success in the gaming market is because it grows from within the needs of play, while generative AI is deployed primarily as a cost-reduction tool. This difference is further solidified through AI-disclosure statements that complicate players’ perceptions of how these technologies are utilized in game development. Through the lens of human-centered AI, we suggest that game developers reorient their design focus to what players need and what that AI can uniquely provide to meet those needs.
\begin{acks}
The authors would like to thank Kaylah Facey, Fiona Shyne, and Kutub Gandhi for assistance coding the reviews. We also want to thank Isabelle Lundin at the Northeastern University Writing Center, who helped us with proofreading and writing of this work. Support was provided by Research Computing at Northeastern University (\url{https://rc.northeastern.edu/}) through the use of the Discovery Cluster.
\end{acks}

\bibliographystyle{ACM-Reference-Format}
\bibliography{refs}

\end{document}